\documentclass[aps,letterpaper,twocolumn,prl,footinbib,floatfix,showpacs]{revtex4}
\usepackage{amsmath}
\usepackage{amsfonts}
\usepackage{amssymb}
\usepackage[scanall]{psfrag}
\usepackage{psfrag}
\usepackage{graphicx}
\usepackage{color}

\begin{document}
\newcommand{\of}[1]{\left( #1 \right)}
\newcommand{\sqof}[1]{\left[ #1 \right]}
\newcommand{\abs}[1]{\left| #1 \right|}
\newcommand{\avg}[1]{\left< #1 \right>}
\newcommand{\cuof}[1]{\left \{ #1 \right \} }
\newcommand{\bra}[1]{\left < #1 \right | }
\newcommand{\ket}[1]{\left | #1 \right > }
\newcommand{\pil}{\frac{\pi}{L}}
\newcommand{\bx}{\mathbf{x}}
\newcommand{\by}{\mathbf{y}}
\newcommand{\bk}{\mathbf{k}}
\newcommand{\bp}{\mathbf{p}}
\newcommand{\bl}{\mathbf{l}}
\newcommand{\bq}{\mathbf{q}}
\newcommand{\bs}{\mathbf{s}}
\newcommand{\psibar}{\overline{\psi}}
\newcommand{\svec}{\overrightarrow{\sigma}}
\newcommand{\dvec}{\overrightarrow{\partial}}
\newcommand{\bA}{\mathbf{A}}
\newcommand{\bdelta}{\mathbf{\delta}}
\newcommand{\bK}{\mathbf{K}}
\newcommand{\bQ}{\mathbf{Q}}
\newcommand{\bG}{\mathbf{G}}
\newcommand{\bw}{\mathbf{w}}
\newcommand{\bL}{\mathbf{L}}
\newcommand{\ohat}{\widehat{O}}
\newcommand{\up}{\uparrow}
\newcommand{\down}{\downarrow}
\newcommand{\MM}{\mathcal{M}}
\author{Eliot Kapit}
\email{eliot.kapit@physics.ox.ac.uk}
\author{Steven H. Simon}
\affiliation{Rudolf Peierls Center for Theoretical Physics, Oxford University}
\title{3- and 4-body Interactions from 2-body interactions in Spin Models: A Route to Abelian and Non-Abelian Fractional Chern Insulators}
\pacs{71.10.Pm,73.43.-f,05.30.Pr}

\begin{abstract}

We describe a method for engineering local $k+1$-body interactions ($k=1,2,3$) from two-body couplings in spin-$\frac{1}{2}$ systems. When implemented in certain systems with a flat single-particle band with a unit Chern number, the resulting many-body ground states are fractional Chern insulators which exhibit abelian and non-abelian anyon excitations. The most complex of these, with $k=3$, has Fibonacci anyon excitations; our system is thus capable of universal topological quantum computation. We then demonstrate that an appropriately tuned circuit of qubits could faithfully replicate this model up to small corrections, and further, we describe the process by which one might create and manipulate non-abelian vortices in these circuits, allowing for direct control of the system's quantum information content.


 
\end{abstract}

\maketitle

The search for non-abelian anyons has become one of the most important developments in quantum condensed matter physics. Non-abelian anyons are exotic collective modes of gapped topological quantum systems, defined by the unique property that when a pair of identical anyons are adiabatically exchanged, the system's wavefunction rotates between different degenerate states \cite{mooreread,nayaksimon,readrezayi2,fukane,levinwen,thomalegreiter,scharfenberger,kitaev,kitaev2003,cooper,barkeshliqi}. These states cannot be distinguished by local operations, and since the creation or destruction of anyons is suppressed by an energy gap, quantum information encoded with anyons will be topologically protected against noise. So far, the most promising systems \cite{willett,anbraiding,mourik} for observing non-abelian anyons are superconducting heterostructures and the 2d electron gas at filling $\nu = 5/2$, and numerous other systems exist in the literature.

Arguing for the existence of non-abelian anyons is often extremely difficult. While Majorana zero modes can be derived straightforwardly at a mean-field level in superconducting heterostructures \cite{fukane}, most other realistic models are not amenable to standard techniques such as perturbation theory or quantum Monte Carlo. Much (though not all) of our theoretical justification for non-abelian anyon states comes from either small-system numerical studies or from considering Hamiltonians with $k+1$-body interactions (with $k>1$) instead of ordinary 2-body interactions, which are rarely realistic for physical systems. 

We here propose a new set of models which can simulate $k+1$-body interactions through realistic 2-body interactions and thus have non-abelian anyon ground states. These models are constructed by modifying the lattice in models with complex hopping and a flat Chern band \cite{kapitmueller,qi1,regnaultbernevig,sungu,neupertsantos,tangmei} so that each site is replaced by a vertex containing a cluster of $k$ spin-$\frac{1}{2}$ degrees of freedom, with tuned interactions between the spins at each vertex, but not between spins at different vertices. The low-energy manifold of the resulting lattice mimics that of particles hopping on a lattice with a single site per vertex and a hard core local $k+1$-body interaction. The ground states of these models \cite{wangyao,wubernevig,bernevigregnault,liu} are generalizations of the Read-Rezayi wavefunctions \cite{readrezayi2}, the most complex of which has a Fibonacci anyon ground state which is capable of universal topological quantum computation. We will first define three lattice Hamiltonians, and then prove that the ground states of these Hamiltonians are Read-Rezayi states with abelian, Ising, and Fibonacci anyons, respectively \footnote{The abelian, Ising and Fibonacci anyon content corresponds to $k=1,2,3$ in the ${\rm{SU}} \of{2}_{k}$ conformal field theories which describe the ground states of our Hamiltonians.}. These Hamiltonians are charged spin models, which could be realized with arrays of coupled qubits. We propose one such realization at the end of this work.


\begin{center}
\textbf{The Model}
\end{center}

Consider a 2D lattice composed of vertices with $k=1,2$ or $3$ sites clustered closely around each vertex. We denote each vertex by a complex position $z_{n}=x_{n} + i y_{n}$, with the vertex positions and inter-vertex couplings defined by a set of tunnel couplings $J \of{z_{n}, z_{m}}$. We label the sites at that vertex by $p$ running from 1 to $k$, and at each site, we place a spin $\frac{1}{2}$ degree of freedom which we quantize along $\sigma_{n,p}^{z}$. We then define the Hamiltonian
\begin{eqnarray}\label{hams}
H^{(k)} &=& H_{J}^{(k)} + H_{V}^{(k)} - \mu \sum_{n}\sum_{p=1}^{k} \sigma_{n,p}^{z}, \\
H_{J}^{(k)} &=& -\sum_{n \neq m} \sum_{p=1}^{k} \sum_{q=1}^{k} \sqrt{p \; q} \left( J \of{z_{n} , z_{m}} \sigma_{n,p}^{+} \sigma_{m,q}^{-} + H. C. \right) \nonumber
\end{eqnarray}
Here, $ H_{V}^{(k)} = V \sum_n f^{(k)} \of{ \sigma_{n,1}^{z} ... \sigma_{n,k}^{z} }$, where $f^{(k)}$ is a polynomial in the $\sigma_{n,k}^{z}$ matrices at a given site $n$ with a maximum degree of two, meaning that it is composed of only one- and two-spin terms. We note that the hopping matrix elements of $H_{J}^{(k)}$ depend on the spin indices $p$ and $q$ of each vertex only through the amplitude factor $\sqrt{p \; q}$; the complex phases of $H_{J}^{(k)}$ are independent of which spins are being exchanged between a pair of vertices.

The lattice defined by $H_{J}$ is chosen so that $H_{J}^{(1)}$ (with only a single spin per vertex) has a flat single-particle ground state band with unit Chern number. There are many models which satisfy this requirement \cite{kapitmueller,qi1,regnaultbernevig,sungu,neupertsantos,tangmei}; in particular, in a previous work \cite{kapitmueller} one of us demonstrated a choice of $J \of{z_{n},z_{m}}$ which leads to an exactly flat lowest Landau level. If that Hamiltonian is chosen for $H_{J}$ all of our following arguments concerning Read-Rezayi ground states are exact and the ground state and quasihole wavefunctions are known analytically, but the same physics will arise in many of the other lattice flat band models as well. From now on, we let $H_{CB} \equiv H_{J}^{(1)}$ refer to the single particle Chern band part of our Hamiltonian. We note also an alternative spin Hamiltonian \cite{greiterschroeter} which stabilizes Read-Rezayi states through ring exchange terms.

\begin{center}
\textbf{The Read-Rezayi Hamiltonian}
\end{center}

Our goal is to replicate the lattice Hamiltonian \cite{wangyao,wubernevig,bernevigregnault,liu}
\begin{eqnarray}\label{HK}
H_{RR}^{(k)} = H_{CB} + \frac{U}{\of{k+1}!} \sum_{j} \prod_{m=1}^{k+1} \of{n_{j} + 1 - m}. 
\end{eqnarray}
Here, $n_{j} = a_{j}^{\dagger} a_j$ is the number of bosons at site $j$. In this Hamiltonian, there is only a single site per vertex, which can be multiply occupied. The interaction in $H_{RR}^{(k)}$ introduces an energy cost $U$ whenever $k+1$ particles occupy the same site, and is the lattice equivalent to the $k+1$-body $\delta$-function in Read and Rezayi's original work \cite{readrezayi2,greiterwen}. This Hamiltonian has a unique (up to topological degeneracies from the boundary conditions) ground state at $\nu \equiv N/N_{\phi} = k/2$; any states with fewer than $N = k N_{\phi}/2$ particles are degenerate and gapless, and any states with more than $k N_{\phi}/2$ particles must either allow for $k+1$-body occupancy or promote particles into the excited bands, in both cases costing finite energy. From now on, we will consider the limit of $U \to \infty$, as it is this limit which we will replicate through $H_{V}^{(k)}$.


\begin{center}
$k=1$: \textbf{Laughlin}
\end{center}
Let $H_{V}^{(1)} = 0$. The $k=1$ Read-Rezayi state is nothing more than a bosonic Laughlin state at $\nu = 1/2$, the boson equivalent to the $\nu = 1/3$ state of fermions. Its fundamental excitations are abelian anyons with charge $1/2$ and a phase of $\pm \pi/2$ whenever two anyons are exchanged, and it is the simplest state to realize in our model. Showing that $H^{(1)}$ is equivalent to $H_{RR}^{(1)}$ is trivial; since we have only a single spin $\frac{1}{2}$ degree of freedom, if we again identify spin $\uparrow$ with a particle state and $\downarrow$ with an empty site, then $H^{(1)}$ is simply the lattice Chern band Hamiltonian with a local hard-core 2-body interaction between bosons. As the Laughlin state both saturates the Chern band and perfectly screens this interaction, it is the exact ground state for $N = N_{\phi}/2$, and is unique up to topological degeneracies from the geometry of the full system.


\begin{center}
$k=2$: \textbf{Moore-Read and Ising Anyons}
\end{center}

For $k=2$ the relevant Read-Rezayi state is a non-abelian anyon state, with half-vortex excitations with braid statistics equivalent to those of Ising anyons \cite{nayaksimon,mooreread,greiterwen,readrezayi1,nayakwilczek,fradkinnayak}. To stabilize it as a unique ground state, we choose:
\begin{eqnarray}\label{HV2}
H_{V}^{(2)} = V \sum_{n} \of{-\sigma_{n,1}^{z} \sigma_{n,2}^{z} - \sigma_{n,1}^{z} + \sigma_{n,2}^{z}     },
\end{eqnarray}
To demonstrate that $H^{(2)} = H_{RR}^{(2)}$, we consider the full $\sigma^{z}$ basis of spin states at each vertex, $\ket{s_{1} s_{2}}$. Since $H_{V}^{(2)}$ contains only $\sigma^{z}$ operators, these states are all eigenstates of $H_{V}^{(2)}$, with energies:
\begin{eqnarray}
E_{\down \down} = E_{\up \down} = E_{\up \up}, \; E_{\down \up} - E_{\down \down} = +4 V.
\end{eqnarray}
Thus, the action of $H_{V}^{(2)}$ constrains the local basis at each vertex by making $\ket{\down \up}$ energetically expensive where all other configurations are isoenergetic. We now make the identification:
\begin{eqnarray}\label{sigtoa}
\ket{\down \down} \to \ket{0}, \ket{\up \down} \to \ket{1}, \ket{\up \up} \to \ket{2}, \ket{\down \up} \to \ket{b} \\
\sum_{p=1}^{2} \sqrt{p} \;  \sigma_{j p}^{+}  =  a_{j}^{\dagger} \of{1 + \frac{a_{j}^{\dagger} b_{j}}{\sqrt{2}} } + \sqrt{2}   b_{j}^{\dagger}. \nonumber \\
\sum_{p=1}^{2} \sqrt{p} \; \sigma_{j p}^{-} = \of{1 + \frac{b_{j}^{\dagger} a_j}{\sqrt{2}} } a_{j} + \sqrt{2} b_j
\end{eqnarray}
These equations describe a model with two species of bosons, $a$ and $b$, with the constraint that any site can contain up to two $a$ bosons or a single $b$ boson, but not a mix of the two (operators such as $b_{j}^\dagger a_{j}^{\dagger}$ always return zero). The hopping matrix elements for the $b$ bosons scale as $\sqrt{2}J$ instead of $J$. In the large $V$ limit the $b$ bosons vanish from the low-energy theory, and to first order in $\of{J/V}$, $H^{(2)}$ becomes
\begin{eqnarray}\label{H2}
H^{(2)} &=& H_{CB} + \lim_{U \to \infty} U \sum_{i} \prod_{m=1}^{3} \of{n_{i} + 1 - m} \\
& &- \frac{1}{4 V} \sum_{ijk} J_{ij} J_{jk} \of{2-n_j} \of{1-n_j} a_{i}^{\dagger} a_{k}. \nonumber
\end{eqnarray}
$H^{(2)}$ is thus identical to the lattice Read-Rezayi Hamiltonian with $k=2$ in the limit $V \to \infty$. The finite $V$ corrections create an irrelevant correction to the hopping matrix $H_{J}$ and a pair of interaction terms with opposite signs (whose contributions will thus interfere to some degree), all of which are suppressed by $J/4V$. As it has a gap with an energy scale set by $J$, we thus expect that the Read-Rezayi state should be robust against these corrections, provided that $J/V$ is not too large.

\begin{center}
$k=3$: \textbf{Fibonacci, and Anyon Braiding}
\end{center}
To stabilize the $k=3$ state, we must engineer an effective four-body interaction. We choose:
\begin{eqnarray}
H_{V}^{(3)} =  V \sum_{n} \left[  -\of{c_{2} + c_{3} + g} \sigma_{n,1}^{z} \sigma_{n,2}^{z}   + c_{2} \sigma_{n,2}^{z} + c_{3} \sigma_{n,3}^{z} \right.  \nonumber \\
  \left. + g \sigma_{n,1}^{z} \sigma_{n,3}^{z}  - \of{c_{2} + c_{3}} \sigma_{n,1}^{z} -  \of{c_{3} + g}  \sigma_{n,2}^{z} \sigma_{n,3}^{z} \right].
\end{eqnarray}
The proof that $H^{(3)}$ replicates the corresponding Read-Rezayi Hamiltonian has the same structure as the proof for $H^{(2)}$ in the previous section. In the local basis $\ket{s_1 s_2 s_3}$ of a given vertex, we identify:
\begin{eqnarray}
\ket{\down \down \down} \to \ket{0}, \; \ket{\up \down \down} \to \ket{1}, \; \ket{\up \up \down} \to \ket{2}, \; \ket{\up \up \up} \to \ket{3}, \\
\ket{\down \up \down} \to \ket{b}, \; \ket{\down \down \up} \to \ket{c}, \; \ket{\up \down \up} \to \ket{a c}, \; \ket{\down \up \up} \to \ket{d}. \nonumber 
\end{eqnarray}
The ``in order" $a$ states are all isoenergetic, for any choice of $c_{2}$, $c_{3}$ and $g$ in (\ref{hams}). If we choose the energy of these states to be zero, the other states have energy:
\begin{eqnarray}
E_{b} &=& 4 V \of{c_2 +  c_3 + g}, \; E_{c} = 4 c_3 V, \nonumber \\
E_{ac} &=& E_{c} + 4 g V, \; E_{d} = E_{b} - 4 g V. 
\end{eqnarray}
The hopping terms can be treated by notational change which is similar to (\ref{sigtoa}); $H^{(3)}$ thus has four particle species, $a,b,c$ and $d$. The $a$ particles experience a hard-core four-body interaction with each other, hard-core repulsive interactions with the $b$ and $d$ bosons, and a repulsive interaction of strength $4 g V$ with the $c$ bosons. The $b,c$ and $d$ bosons have hard-core repulsive interactions with themselves and each other, and while the $b$ and $c$ bosons can move freely to neighboring empty sites, the $d$ bosons can only move through second order tunneling processes with an amplitude proportional to $J^{2}/V$. They are thus essentially immobile, analogous to the repulsively bound pairs observed in optical lattice experiments \cite{winkler}. For $c_{2}, c_{3}$ and $g$ all positive, these states are all far higher in energy than the $a$ states and are irrelevant to the low-energy physics, so $H^{(3)}$ becomes an exact replica of $H_{RR}^{(3)}$ in the infinite-$V$ limit.

Such a model is more complex, and would undoubtedly be difficult to create in a real array of coupled qubits. However, the consequences of realizing $H^{(3)}$ would be profound: the fundamental excitations of the $k=3$ Read-Rezayi state, called $Z_{3}$ parafermions, are equivalent from a quantum information perspective to Fibonacci anyons, which, unlike Majorana modes, are capable of universal quantum computation. In other words, any desired unitary transformation on the system's non-abelian degenerate subspace can be generated through a suitable number of braids, so quantum algorithms can be executed with every operation protected by the system's energy gap.

If we choose a different set of parameters in $H^{(3)}$ we can obtain a model of spin-1/2 bosons with a hard-core three-body interaction such that a site may be occupied by two or fewer bosons of any combination of spins at no additional energy cost but triple occupancy is forbidden. The precise details of this implementation are described in the supplemental information to this paper; the spin degree of freedom is conserved by choosing the energies of the two species to be widely separated so that the conversion of an excitation from one type to the other is forbidden by energy conservation. The lattice could be populated by applying a bichromatic drive signal, as described below. This model is gapless for $\nu < 4/3$, and at $\nu = 4/3$ the exact ground state is a non-abelian spin singlet state \cite{ardonneschoutens}, which also has Fibonacci anyon excitations. It is thus an alternative to the $k=3$ state of a single species of bosons described above; both states support universal topological quantum computation.

Further, in all three of these models, the fundamental quasihole excitations can be bound and manipulated through simple external operations. In any Read-Rezayi state, the fundamental quasiholes are vortices which cause the wavefunction to vanish as $k$ particles approach their location. If we modify the Hamiltonian by locally adding one or more potential terms $+ U_{n} \sigma_{n k}^{z}$ acting on the $k$-th spin at a given vertex $n$, we can insert local $k-$body contact repulsive impurities. If the energy $U_{n}$ is sufficiently large, or if additional flux quanta are inserted into the system, then fractional quasiholes will be pinned at the impurity sites, as they exactly eliminate the energy cost of the impurities. By adiabatically moving these potentials around the lattice, non-abelian quasiholes can be braided to perform quantum logic gates \cite{nayaksimon,nayakwilczek,georgiev,freedmannayak,tserkovnyaksimon,barabanzikos,prodan,kapitbraiding}.

\begin{center}
\textbf{Qubit Implementations and Loss Processes}
\end{center}

The most difficult aspects of engineering (\ref{hams}) in a real qubit array would undoubtedly be maintaining local addressibility and device parameter homogeneity across a large lattice and generating the artificial gauge field which makes $J$ complex. While addressing the former depends on the specific details of the physical qubits and is beyond the scope of this discussion, in a recent work, one of us showed that artificial gauge fields of arbitrary magnitude and complexity can be engineered in a qubit lattice \cite{kapitgauge}. Briefly, the proposal consists of a decorated lattice of two types of qubit, labelled by $A$ and $B$. We choose device parameters so that the $B$ qubits are higher energy than the $A$ qubits, so that we can integrate them out to generate mediated hopping terms between $A$ qubits. The couplings between pairs of qubits are either purely ``$\pm$" ($\sigma_{A}^{+} \sigma_{B}^{-} + \sigma_{A}^{-} \sigma_{B}^{+}$ in the qubit basis, or parity-odd operators in the local eigenbasis of each physical qubit device) or ``$zz$" ($\sigma_{A}^{z} \sigma_{B}^{z}$, or parity-even operators) couplings. We then apply an oscillating field $\widehat{V} \sin \omega t + \varphi_s$ to all the qubits and allow the phase offset $\varphi_s$ to vary from site to site. This field breaks time reversal symmetry, as it becomes impossible to choose an origin for the time coordinate such that $H \of{t} = H \of{-t}$ if $\varphi_s$ varies spatially. In a frame rotating at frequency $\omega$, this phase offset acts like a unitary rotation on the Bloch sphere of each qubit, and the tunneling matrix elements for the rotating frame excitations generated by the anisotropic $\pm$ and $zz$ couplings are complex with differing dependence on the local phase offsets. The phase accumulated around a closed path on the lattice which includes sites with multiple values of $\varphi_s$ and both types of coupling can therefore be nonzero, realizing a nontrivial artificial gauge field.


To realize $H^{(k)}$, we would place $k$ $A$ qubits at each vertex, and couple each to a single $B$ qubit per link, with the magnitudes of the $A-B$ couplings scaled by the appropriate factors of $\sqrt{p}$. We would then implement $H_{V}^{(k)}$ through appropriately tuned couplings and add weak single qubit energy shifts and weak exchange terms to cancel the intra-vertex tunneling matrix elements generated by the coupling to the $B$'s. The level of fine-tuning required to achieve this is significant, but since all unwanted terms are generated by well-controlled small parameter expansions, corrections to $H^{(k)}$ can be systematically eliminated using appropriate additional couplings. Note that in any driven system implementation of the model described here, $\sigma^{z}$ should be chosen to represent a potential term in the local (rotating frame) occupation basis of each physical qubit. 


Given a lattice realization of our Hamiltonian, we must next consider how to prepare the system in its topological ground state. We shall assume that there are no spontaneous processes which flip a spin from $\down$ (or the ``empty" state in the rotating frame) to $\up$ (the ``particle" state), and that the $\up$ spins can decay back to $\down$ with a rate $\Gamma_D$, where $\Gamma_D \ll J$. If the default energy cost to create an $\up$ particle is $\omega_0$, then below $N < N_\phi k/2$ the minimum cost to add a particle is just $\omega_0$ but for any $N \geq N_\phi k/2$ the minimum cost to add a particle is $\omega_0 + \Delta$, where $\Delta$ is the system's many-body gap. To populate the system, we can thus apply a uniform drive field at frequency $\omega_0 < \omega < \omega_0 + \Delta$; in the rotating frame, the drive field will shift the chemical potential by $+ \omega$ and introduce a mixing term $\sum_i \Omega \sigma_{i}^{x}$. If both $\Gamma_D$ and $\Omega$ are small compared to $\Delta$, these terms cannot close the bulk gap and the system's ground state will be the Read-Rezayi state of level $k$. 

It is important to note that, in any physical qubit implementations of the $k=2$ and $k=3$ models, small populations of the higher energy boson species will be generated by decay processes. So long as the rates of these processes are slow enough that the system can remain in equilibrium and retain its energy gap, the addition of the other high-energy boson species will not disrupt the non-abelian information content of the system except possibly after very long times \footnote{The only way for a loss process to change the topological charge in a region is for the created quasiholes and quasiparticles to fractionalize and then separate widely.}. This is because the $b$ (and $c,d$ for $k=3$) bosons have a hard-core interaction with the $a$ bosons, and therefore will nucleate full quasiholes in the $a$ boson wavefunction to completely screen this interaction. Such quasiholes are topologically trivial aside from abelian exchange phases, and cannot change the non-abelian anyon content of the $a$ boson wavefunction. Even if we were to increase the $b/c$ boson density substantially, they would at most form one or more abelian ``221" states alongside the higher order $a$ boson state, which will not alter the non-abelian part of the fractional statistics of the $a$ boson excitations. 

The system can be sympathetically cooled by weakly coupling it to one or more reservoir arrays of qubits which are tuned to only accept excitations with energies near $\omega_0 + \Delta$ or $E_{b/c/d}$. By repeatedly measuring the state of the reservoir array and applying NOT operations to remove any excitations, we can maintain it at effectively zero temperature and thus provide a continuous cooling mechanism for the primary spin lattice. Such sympathetic cooling would likely be necessary to preserve the topological ground state for long times.

\begin{center}
\textbf{Conclusion}
\end{center}

We have proposed a set of qubit arrays for which we can analytically demonstrate the existence of abelian and non-abelian ground states. Further, the most ambitious formulation of our model, with three qubits per lattice vertex, supports a Fibonacci anyon ground state, and thus could be used for universal quantum computation. Due to the ground state energy gap, these systems are resistant to disorder and local noise, though quantifying this resistance and understanding its limits will require many-body numerical simulations which are beyond the scope of this letter.


\begin{center}
\textbf{Acknowledgements}
\end{center}

We would like to thank M. Hafezi, E. J. Bergholtz and Z. Liu for helpful discussions. This material is based on work supported by EPSRC Grant Nos. EP/I032487/1 and EP/I031014/1, and Oxford University.

\bibliography{FQgauge}

\begin{center} \textbf{Supplemental Information} \end{center}

\begin{center}
\textbf{Spin $\frac{1}{2}$ Bosons in a Qubit Array}
\end{center}
We can add a pseudospin degree of freedom to our array through a modification of $H^{(3)}$. Consider a lattice with 3 physical spins per vertex, and the Hamiltonian
\begin{eqnarray}
H_{S}^{(2)} &=& -\sum_{n \neq m} \sum_{p,q=1}^{k} f_p f_q   \left( J \of{z_{n} - z_{m}} \sigma_{n,p}^{+} \sigma_{m,q}^{-} + H. C. \right) \nonumber \\
& &+ \frac{1}{2} \sum_{n} \left[   \omega_{0} \sigma_{n,1}^{z} + \frac{2 \omega_0 -3 \omega_{d} + 3 \omega_{u}}{2} \sigma_{n,2}^{z} \right.  \\
& & \left. + \frac{\omega_{d} + \omega_{u}}{2} \sigma_{n 3}^{z} + \of{\omega_0 - \omega_d} \sigma_{n,1}^{z} \sigma_{n,3}^{z} \right. \nonumber \\
& & \left. + \frac{2 \omega_0 - 3 \omega_d + \omega_u}{2} \sigma_{n,2}^{z} \sigma_{n,3}^{z} \right], \nonumber
\end{eqnarray}
where $f_p = \sqrt{2}$ if $p=3$ and 1 otherwise. We now make the following state identifications
\begin{eqnarray}
\ket{\down \down \down} \to \ket{0}, \; \ket{\up \down \down} \to \ket{d}, \; \ket{\up \up \down} \to \ket{u d}, \;  \\
\ket{\down \up \down} \to \ket{u}, \;  \ket{\up \down \up} \to \ket{2d}, \; \ket{\down \up \up} \to \ket{2u}. \nonumber \\
\ket{\down \down \up} \to \ket{\psi_{1}}, \;\ket{\up \up \up} \to \ket{\psi_{2}}. \nonumber
\end{eqnarray}
The energies of these states are
\begin{eqnarray}
E \of{n_u, n_d} &=& \omega_u n_u + \omega_d n_d, \\
 E \of{\psi_1 } &=& 3 \omega_d - 2 \omega_0, \; E \of{\psi_{2}} = 2 \omega_u - \omega_d +2 \omega_0. \nonumber
\end{eqnarray}
Let us choose $\omega_u, \omega_d$ and $\omega_0$ so that $\abs{J/\of{\omega_u - \omega_d}} \ll 1$ and the energies of the two ``error states" $\psi_1$ and $\psi_2$ are far apart from any of the $u$ and $d$ states. In this limit, the populations of the two ``spin" states $u$ and $d$ are fixed by energy conservation, and we obtain an interacting model of two species of lowest Landau level bosons with a hard-core 3-body interaction (for any combination of spins). The $\psi_{1}$ particles are mobile and have a hard-core interaction with the other boson species (analogous to the $b$ bosons in the $k=2$ model), and the $\psi_2$ bosons are immobile ``repulsively bound pairs" (analogous to the $d$ bosons in the single-species $k=3$ model). As mentioned in the text, the ground state of this system at $\nu = 4/3$ is a non-abelian spin singlet state \cite{ardonneschoutens}, which also has Fibonacci anyon excitations. 

We note also that in some physical qubit implementations (such as the one described in section III.C of \cite{kapitgauge}) of an artificial gauge field, one must consider higher excited modes of the individual qubit devices, and they can no longer be regarded as simple spin-1/2 objects. In this limit one would typically choose one or more excited states to be the ``particles" of the theory and generalize the $\sigma^{\pm}$ and $\sigma^{z}$ operators to be the creation/annihilation and potential terms for these states.

\end{document}